\documentclass[twocolumn,pra,twocolumn,showpacs,amsmath,amssymb,preprintnumbers]{revtex4}
\usepackage{hyperref}
\usepackage{graphicx}
\usepackage{dcolumn}
\usepackage{bm}

\usepackage{amsthm}

\newcommand{\prob}[1]{\mathbb{P}_{\mathrm{QM}}\!\left( #1 \right)}
\newcommand{\probL}[1]{\mathbb{P}_{\mathrm{LR}}\!\left( #1 \right)}

\newcommand{\expec}[1]{\left<#1\right>}

\newcommand{\ket}[1]{ | #1 \rangle}
\newcommand{\bra}[1]{ \langle #1 |}

\newcommand{\la}{\lambda}

\def\one{\leavevmode\hbox{\small1\normalsize\kern-.33em1}}

\newcommand{\beq}{\begin{equation}}
\newcommand{\eeq}{\end{equation}}
\newcommand{\bea}{\begin{eqnarray}}
\newcommand{\eea}{\end{eqnarray}}

\newcommand{\lai}{\lambda_i}
\newcommand{\laj}{\lambda_j}
\newcommand{\expecr}[1]{\left<#1\right>_{\mathrm{HS}}}
\newcommand{\expecrr}[1]{\left<#1\right>_{\!\!\! \mathrm{HS}}}
\newcommand{\expecO}[1]{\left<#1\right>_{\!\!\!  O}}

\begin{document}

\title{Violations of Bell inequalities from random pure states}

\author{Max R. Atkin$^1$ and Stefan Zohren$^2$}

\affiliation{$^1$ Institut de Recherche en Math\'{e}matique et Physique, Universit\'e Catholique de Louvain, Belgium \\
$^2$ Quantum and Nanotechnology Theory Group, Department of Materials, and \\
Machine Learning Research Group, Department of Engineering Science, University of Oxford, UK}

\date{May 25, 2015}

\begin{abstract}
We consider the expected violations of Bell inequalities from random pure states. More precisely, we focus on a slightly generalised version of the CGLMP inequality, which concerns Bell experiments of two parties, two measurement options and $N$ outcomes and analyse their expected quantum violations from random pure states for varying $N$, assuming the conjectured optimal measurement operators. It is seen that for small $N$ the Bell inequality is not violated on average, while for larger $N$ it is. Both ensembles of unstructured as well as structured random pure states are considered. Using techniques from random matrix theory this is obtained analytically for small and large $N$ and numerically for intermediate $N$. The results show a beautiful interplay of different aspects of random matrix theory, ranging from the Marchenko-Pastur distribution and fixed-trace ensembles to the $O(n)$ model.
\end{abstract}
\maketitle

\section{Introduction}
In quantum information theory ``non-classical'' properties of a quantum system are of importance for many of its applications \cite{Nielsen}. There are different ways of classifying what is meant by a state of quantum system to be ``non-classical''. Let us focus on a composite system $AB$ made up of two sub-systems $A$ and $B$ which for simplicity we assume to have Hilbert spaces of equal dimension $N$. One quantity classifying the ``non-classical'' correlations between $A$ and $B$ is \emph{entanglement entropy} \cite{Ben96}, measured by the von Neumann entropy of the reduced density matrix of either  sub-system. In the above case that both of the subsystems are $N$-dimensional, its maximum value is $\log N$ and the state with this entropy is referred to as the \emph{maximally entangled state}. Another measure for ``non-classicality'' of a quantum system is given by the violations of Bell inequalities. Besides Bell's original inequality \cite{Bell}, one of the most well known Bell inequalities is the so-called CHSH inequality \cite{CHSH} which concerns a Bell experiment with two party system, Alice and Bob, each having two measurement options which can have two different outcomes. In the case of the CHSH inequality the maximal violation (under optimal measurement operators) is caused by the maximally entangled state, leading to an agreement between both measures of ``non-classicality'  introduced above.

Recently there has been a considerable interest in \emph{random pure states} (see \cite{random-density} and references therein) which for example emerge due to noise in the preparation of the state or when the state is evolving in time under chaotic dynamics. Since random states arise from random density matrices their formulation involves techniques from \emph{random matrix theory} (RMT) (see \cite{Metha} and \cite[Chapter 1]{RMTbook} for an overview). One particularly interesting aspect of (unstructured) random pure states of a bipartite system is that the expected value of its entanglement entropy is given by $\log N (1 - (2\log N)^{-1})$ \cite{Lubkin-Lloyd,Page} (see also \cite{Nadal} for the full distribution function for large $N$), i.e.\ for large $N$ it approaches the maximal value, associated to the maximally entangled stated. In the context of the above discussion it is interesting to consider the violations of Bell inequalities under random pure states. This letter aims to give a first account of such an analysis.

%
%
\section{Bell inequalities and their quantum violations from random pure states}
%
A generalisation of the CHSH inequality in the case of $N$ possible measurement outcomes is given by the CGLMP inequality \cite{CGLMP}. It has a slightly generalised and simplified version \cite{ourBell},
\bea
\probL{A_2<B_2}+\probL{B_2<A_1}+\probL{A_1<B_1}+\nonumber  \\
+\probL{B_1\leq A_2}\geqslant 1,\label{ourBell}
\eea
where $\probL{A_a<B_b}$ denotes the probability, under \emph{local realism} (LR), that the outcome of Alice's measurement, having chosen the measurement option $a=1,2$, is less than the outcome of Bob's measurement, having chosen the measurement option $b=1,2$; both $A_a$ and $B_b$ taking values in the same set of $N$ possible outcomes.

In the case of quantum mechanics (QM), as opposed to local realism, the corresponding probabilities are given by
\bea\label{eq:QM}
\prob{A_a<B_b}=\sum_{k<l} \mathrm{tr}\left(\rho \,A_a^k \otimes B_b^l \right),
\eea
where $\rho$ is the density matrix of the bipartite system $AB$, with Hilbert space $\mathcal{H}_A\otimes \mathcal{H}_B$, while $A_a^k$ and $B_b^l$ are positive operators on the sub-systems $A$ and $B$ with $\sum_{k=1}^N A_a^k= \mathbb{I}$ and $\sum_{l=1}^N B_b^l=  \mathbb{I}$; here the indices $a,b$ label the two possible measurement options and $k,l$ the $N$ possible outcomes. 

It is well known that quantum mechanics can violate the above Bell inequality which occurs when 
\bea
\mathcal{A}_N&=&\prob{A_2<B_2}+\prob{B_2<A_1}+\nonumber \\
&+&\prob{A_1<B_1}+\prob{B_1\leq A_2}
\eea
is less than one and the maximal violation corresponds to the minimal value of $\mathcal{A}_N$, which we often refer to as the { \em{minimal target value}}. The state that achieves the minimal target value is referred to as the {\em{optimal state}}. Quantum violations of the above Bell inequality and the original CGLMP inequality were studied numerically for $N>2$ \cite{ourBell,optimal} (see also \cite{ourquantumBell} for an exact result for $N\to\infty$), where it was seen that the maximal violation is obtained for a pure state under earlier conjectured best measurement operators \cite{CGLMP,Zukowski}. It was observed for $N>2$ that although the optimal state is not the maximally entangled state, the optimal measurement operators remain the same for both states. Let us therefore assume that the operators $A_a^k$ and $B_b^l$ are given by the conjectured best measurement operators and that $\rho=\ket{\psi}\bra{\psi}$ is pure. It is useful to introduce the Schmidt decomposition of the pure state
\beq
\ket{\psi}=\sum_{i=1}^N \sqrt{\la_i} \,\, \ket{i}_A\otimes \ket{i}_B, \label{Schmidt}
\eeq
where the so-called Schmidt coefficients $\{\la_i\}$ are non-negative and normalised to $\sum_i \la_i=1$. Recall that the maximal entangled state, here denoted by $\ket{\Phi}$ has $\la_i=1/N$ for all $i=1,...,N$, i.e.\
\beq
\ket{\Phi}=\frac{1}{\sqrt{N}} \sum_{i=1}^N \ \ket{i}_A\otimes \ket{i}_B.
\eeq

\begin{figure}[t]
\begin{center}
\includegraphics[width=2.5in]{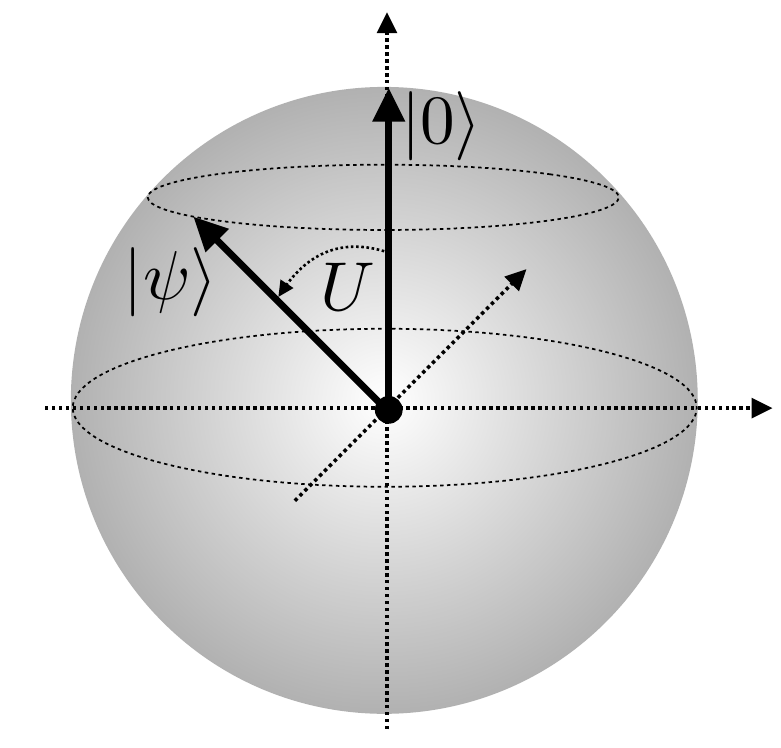}
\caption{Illustration of the Bloch sphere. Any pure state in $\mathcal{H}$ corresponds to a point on the sphere, while mixed states correspond to points within the unit ball. A random pure state of the Hilbert-Schmidt ensemble is obtained by choosing a point on the sphere uniformly at random. This can be done by applying a random rotation $U$ to the references state $\ket{0}$, here chosen to be the north-pole.}
\label{fig0}
\end{center}
\end{figure}

It was shown in \cite{ourBell} that under the above assumptions one has
\beq\label{target}
\mathcal{A}_N(\{\la_i\})= \sum_{i,j=1}^N M_{ij} \sqrt{\lai\laj}, 
\eeq
with $M_{ij} =2 \delta_{ij} - \frac{1}{N} P_{ij}$ and 
\beq
 P^{(N)}_{ij}=\sec\left(\frac{(i-j)\pi}{2 N}\right),
\eeq
where $\sec(\cdot)=1/\cos(\cdot)$ is the secant.

Our aim is to analyse the quantum violations of the above Bell inequalities \eqref{ourBell} for random pure states. Let us now summarise the main results which are derived in the following section. Assuming that the best measurements are the same as for the maximally entangled and optimal state, we can analyse the quantum violations of the Bell inequality \eqref{ourBell} by means of expression \eqref{target}. The above assumption is justified by the extensive numerical analysis presented in \cite{ourBell}, where the optimal measurements were obtained numerically for both the maximally entangled state, as well as when optimised jointly over states. Furthermore, since the ensembles of random states considered here are invariant under permutations of the eigenvalues, it is expected that the best measurements are the same as for the maximally entangled state. 

The most natural definition of a random pure state is to take a point on the higher-dimensional analog of the Bloch sphere uniformly at random as illustrated in Figure \ref{fig0}. Recall that any pure state $\ket{\psi}\in \mathcal{H}_A\otimes \mathcal{H}_B$ can be represented by a point on the space $\mathcal{U}(N^2)/(\mathcal{U}(N^2-1)\times \mathcal{U}(1))$ which is the higher-dimensional analog of the Bloch sphere for $\mathcal{H}=\mathcal{H}_A\otimes \mathcal{H}_B$. The measure of such random pure states is often also called the \emph{Hilbert-Schmidt} (HS) ensemble. Following the exposition in \cite{random-density}, we can fix an arbitrary reference point $\ket{0}_{AB}=\ket{0}_A\otimes \ket{0}_B$, i.e.\ the north-pole on the Bloch sphere, and represent a random pure state from the HS ensemble by a random rotation of this state,
\beq \label{defHS}
\ket{\psi}_{\mathrm{HS}} = U_{AB} \ket{0}_{AB},
\eeq
where $U_{AB}$ is chosen uniformly according to the Haar measure on the set $\mathcal{U}(N^2)$ of unitary matrices on $\mathcal{H}$.

\begin{figure}
\begin{center}
\includegraphics[width=3.1in]{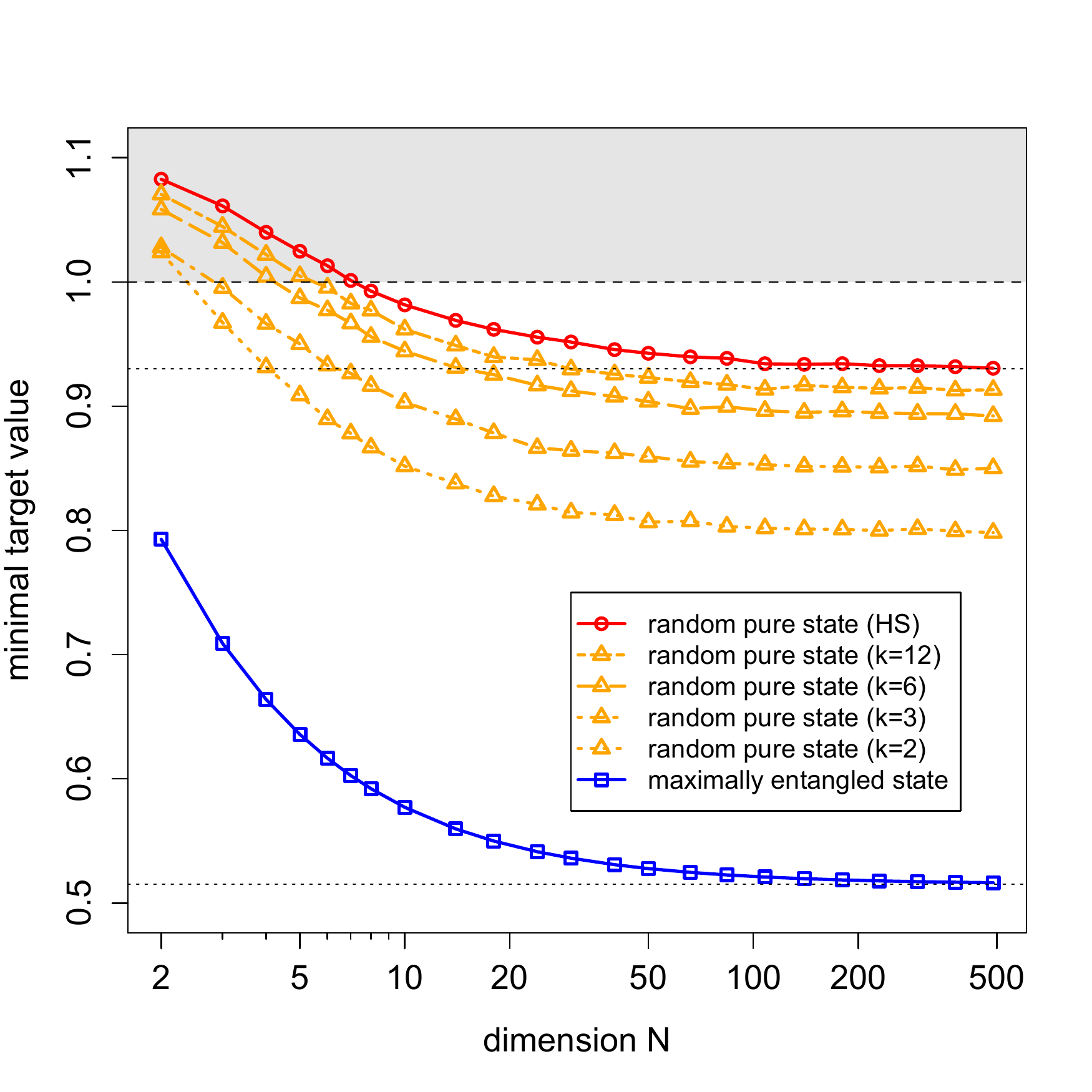}
\caption{Mean minimal value of $\mathcal{A}_N$ under the measure of random pure states from the HS ensemble (red circles) in comparison to the corresponding value for a maximally entangled state (blue squares), both as a function of $N$. The dotted lines correspond to their asymptotic values. In addition, the mean minimal value of $\mathcal{A}_N$ under the measure of random pure states from the structured ensemble (orange triangles) is plotted as a function of $N$ for $k=12,6,3,2$ (from top to bottom). Indicated in grey is the region where the Bell inequality is not violated.}
\label{fig1}
\end{center}
\end{figure}

As we will see in the next section for a random pure state from the HS ensemble the Schmidt coefficients $\{\lai\}$ in \eqref{Schmidt} and \eqref{target} can be related to the square singular values of random \emph{Wishart matrices} which are constrained to have $\sum_i \lai=1$. 
By using techniques from RMT we are able to calculate the expected value $\expecr{\mathcal{A}_N}$ of \eqref{target}, where the expectation value is taken with respect to the measure of the HS ensemble. This is done analytically for both $N=2$ and large $N$, while  for intermediate values of $N$, we compute $\expecr{\mathcal{A}_N}$ numerically. The results are summarised in Figure \ref{fig1}. Shown is the expected minimal target value $\expecr{\mathcal{A}_N}$ under a random pure state in comparison to the minimal target value $\mathcal{A}_N$ for the maximally entangled state, for $N$ varying from 2 to 500. The numerical data shows that the maximally entangled state always violates the Bell inequality \cite{ourBell}, while for the random pure state one observes that the expected target value falls below $1$ for $N \gtrsim 8$. We note that the variance of $\mathcal{A}_N$ under the measure of the HS ensemble goes to zero as $N\to\infty$. This means that for large $N$ almost all random states sampled from the HS ensemble  violate the Bell inequality. Further explanation is provided in Appendix \ref{appA} and particularly in Figure \ref{fig2}. 

Using techniques from RMT we obtain analytically the value for $N=2$ given by $\expecr{\mathcal{A}_2} = 3/2 - 3 \pi/(16 \sqrt{2})\approx 1.083$, as well as the asymptotic value as $N\to\infty$ given by $\expecr{\mathcal{A}_\infty} = 2-1024\, G/(9 \pi^4)\approx 0.93$, with $G$ being Catalan's constant. In Appendix \ref{appA} we also indicate how to analytically obtain $\expecr{\mathcal{A}_N}$ at finite $N$ together with the $k$-th moments. 

In the above discussion we considered the violations of the Bell inequality \eqref{ourBell} from random states from the HS ensemble and compared it to the violation from maximally entangled states. One can view both cases as the extremes of structured ensembles as introduced in \cite{random-density} - the HS ensemble being the ``most unstructured'' and the maximally entangled state being the ``most structured''. More precisely, for a fixed $k=1,2,3,...$ the structured ensemble is obtained from a superposition of $k$ maximally entangled states $\Phi$, each rotated in one of the subsystems (say $A$) by a random unitary matrix $U_i$, $i=1,...,k$, chosen uniformly according to the Haar measure on the set $\mathcal{U}(N)$ of unitary matrices on $\mathcal{H}_A$,
\beq \label{structuredef}
\ket{\psi}_{\mathrm{k}} = \left[ \left(\sum_{i=1}^k U_i \right) \otimes  \mathbb{I}_B \right] \ket{\Phi}
\eeq
This ensemble interpolates between the maximally entangled state for $k=1$ and the HS ensemble in the limit $k\to \infty$. We analysed the expected value $\expec{\mathcal{A}_N}_{k}$ under the structured ensemble with parameter $k$. This is done for both analytically for large $N$, as well as numerically for intermediate $N$ and fixed values $k=2,3,6,12$. The results are also shown in Figure \ref{fig1} together with the corresponding values for the HS ensemble ($k\to \infty$) and maximally entangled state ($k=1$). As shown in the next section, using RMT, we obtain the asymptotic value of $\expec{\mathcal{A}_N}_k$ as $N\to\infty$ given by $\expec{\mathcal{A}_\infty}_k \approx  0.796, 0.848, 0.892, 0.912$ for $k=2,3,6,12$ respectively. The full analytical expression for arbitrary $k$ is derived in the following section, cf.\ \eqref{reskens}.

%
%

\section{Random matrix theory and random pure states}

\subsection{Uniform random states and the Hilbert Schmidt ensemble}

We defined a random pure state from the Hilbert-Schmidt (HS) ensemble in \eqref{defHS} through a random unitary
matrix acting on some fixed reference state. This is equivalent to the \emph{Fubini-Study measure}. In \cite{Page} it was shown that
such a random pure state has Schmidt coefficients distributed according to
the \emph{joint probability distribution function} (jpdf),
\beq\label{jpdf}
\mathbb{P}_{\mathrm{HS}}(\{\lai\})=\frac{1}{Z_N} \prod_{i<j}|\lai-\laj|^2\delta\left(\sum_{i=1}^N\lai -1\right)
\eeq
where the \emph{partition function} $Z_N$ is the normalisation such that $\int_0^\infty P(\{\lai\}) \prod_i d\lai=1$. Up to the constraint, this is the eigenvalue distribution of a complex Wishart matrix \cite{Wishart}, i.e.\ real and imaginary parts of all entries are $i.i.d.$ normal random variables. We exploit this fact in the numerical investigation as explained in Appendix \ref{appB}.

One way to arrive at the above expression is to note that the Fubini-Study measure implies
that an (unnormalised) random pure state is given by
\beq
\ket{\psi} = \sum_{i j=1}^N X_{ij} \ket{i}_A\otimes \ket{j}_B,
\eeq
where $X$ is distributed according to
\beq
\mathbb{P}_{\mathrm{HS}}(X)\propto  \delta(\mathrm{tr}(X X^\dag) -1) dX.
\eeq
Here $dX=\prod_{ij} d\Re X_{ij} d\Im X_{ij} $ is the Haar measure. We first note that $M=X X^\dag$ is a Hermitian matrix. Diagonalising $M=V^\dag \Lambda V$ with $\Lambda=\mathrm{diag}(\lambda_1,...,\lambda_N)$ and integrating out the angular degrees of freedom one arrives at an expression identical to \eqref{jpdf} where the $\{\lai\}$ are the eigenvalues of $M$. Here the term $\Delta^2(\Lambda)=\prod_{i<j}|\lai-\laj|^2$ is called the Vandermonde determinant and comes from the Jacobian of the change of variables $M\to (V,\Lambda)$. The last step is now to identify the eigenvalues of $M$ or equivalently the square singular values of $X$ with the Schmidt coefficients. It is  simple to show that the reduced density matrix for either sub-system takes the form,
\beq\label{Ginrho}
\rho_A=\rho_B= \frac{X X^\dag}{\mathrm{tr} (X X^\dag)}.
\eeq
Since $\rho_A=\sum_i \la_i \ket{i}_A \bra{i}_A$ and $\rho_B=\sum_i \la_i \ket{i}_B \bra{i}_B$, we identify the (unnormalised) Schmidt coefficients with the square singular values of $X$ which leads to \eqref{jpdf}.

Our primary interest is in how $\expecr{\mathcal{A}_N}$ varies with $N$. We begin writing down the expectation value of $\mathcal{A}_N$ with respect to the measure \eqref{jpdf}; from \eqref{target} one gets
\beq
\label{avgA}
\expecr{\mathcal{A}_N} = 2 - \frac{1}{N} - \frac{2}{N} \sum_{i< j} P_{ij} \expecr{\sqrt{\lambda_i \lambda_j}}. 
\eeq
We first consider the case $N=2$, which from the above discussion has the integral representation,
\beq
\expecr{\mathcal{A}_2} = \frac{3}{2} - \frac{1}{Z_2} \sum_{i < j}P_{ij} \int^1_0 \!\!\! d^2\! \lambda (\lambda_1 - \lambda_2)^2 \sqrt{\lambda_1 \lambda_2}.
\eeq
where we note that due to the permutation symmetry of the jpdf we may factor the integral out of the sum over $i$ and $j$. Evaluating this expression yields,
\beq\label{A2}
\expecr{\mathcal{A}_2} = \frac{3}{2} - \frac{3 \pi}{16 \sqrt{2}}
\eeq
as announced above. The behaviour of $\expecr{\mathcal{A}_N}$ as $N\rightarrow \infty$ can be found by a saddle point analysis of \eqref{avgA}. Central to this approach is the eigenvalue density $\rho(\lambda,N)=\expecr{\frac{1}{N}\sum_i \delta(\la-\lai)}$ of the jpdf \eqref{jpdf} for large $N$. From the discussion surrounding equation \eqref{Ginrho} one sees that the spectral density is exactly the \emph{Marchenko-Pastur distribution} \cite{Marchenko-Pastur} in the case of square matrices. Another route to obtain the eigenvalue density is to start from the partition function $Z_N$ associated to \eqref{jpdf} and write the constraint as a Lagrange multiplier in the effective action. The spectral density can then be found using a saddle point analysis of the resulting Coulomb gas model. Yet another alternative is to note that the jpdf \eqref{jpdf} is known in the literature as a fixed-trace ensemble and has been studied in the works \cite{Metha,Page,FT,Gernot1,FTothers}. 
By considering a fixed trace $\sum_i\la_i=t$, taking the Laplace transform of expectation values with respect to $t$, rescaling the eigenvalues, transforming back and setting $t=1$, one finds the relation,
\beq
\label{deltaLUE}
\expecrr{\prod^r_{i=1} \lambda_i^{\eta_i}} = \frac{\Gamma(N^2)}{\Gamma(N^2 + \eta)}N^\eta \expec{\prod^r_{i=1} \lambda_i^{\eta_i}}_\mathrm{LUE}
\eeq
where $\eta_i \in \mathbb{R}$, $\eta = \sum^r_{i=1}\eta_i$ and the expectation on the right-hand-side is with respect to the Laguerre ensemble defined by the jpdf,
\beq
\mathbb{P}_\mathrm{LUE}(\{\lambda_i\}) = \frac{1}{Z_\mathrm{LUE}}    \prod_{i<j}|\lai-\laj|^2 e^{-N \sum_i \lambda_i} .
\eeq
Using the well known form of the spectral density of the Laguerre ensemble together with \eqref{deltaLUE} we obtain,
\bea
\rho(\lambda,N) &=& N \mu(N \lambda) \nonumber\\
\mu(x) &=& \frac{\sqrt{4-x}}{2\pi \sqrt{x}}, \quad \text{for $x\in [0,4]$} \label{density}
\eea
and $\mu(x)=0$ otherwise (see also \cite{random-density}). Let us remark that the fact $\rho(\lambda,N)\approx N \mu(N \lambda)$ shows the spacing between eigenvalues is of order $1/N$ as expected.

We now proceed by utilising the invariance of the jpdf under permutations of the eigenvalues which allows \eqref{avgA} to be written as,
\bea
\label{avgA2}
 &&\expecr{\mathcal{A}_N} = 2 - \frac{1}{N} +\nonumber \\
&& \!\!\! -\!\! \left(\frac{2}{N^2} \sum_{i < j} P_{ij}\right)\! \expecrr{\frac{1}{N}\sum_k \frac{1}{N-1}\sum_{l\neq k} N\sqrt{\lambda_k \lambda_l}}\!\! . 
\eea
For $N$ large we make the following approximations,
\beq \label{tmpxx2}
\frac{1}{N^2} \sum_{i < j} P_{ij} \approx \int^1_0dx\int^x_0dy \sec\left(\frac{\pi(x-y)}{2} \right) = \frac{8G}{\pi^2}
\eeq
where $G$ is Catalan's constant and
\bea
\frac{1}{N^{2}}\expecrr{\sum_k \sum_l N\sqrt{\lambda_k \lambda_l}}\!\!\!  &\approx&  \!\! \int_0^\infty \!\!\! \int_0^\infty \!\!\! \!\! \mu(x)\mu(y) \sqrt{xy} dx dy \nonumber \\
&=& \!\!  \frac{64}{9 \pi^2}.
\eea
This leads to the result, already announced above,
\beq\label{Ainf}
\expecr{\mathcal{A}_\infty} = 2-\frac{1024\, G}{9 \pi^4}.
\eeq

\subsection{Random states from structured ensembles}

The definition of a random density matrix from the structured ensemble with parameter $k$ is given in \eqref{structuredef}. It is not hard to see that when taking the partial trace on any of the subsystem $A$ or $B$ the reduced density operator is given by 
\beq\label{krho}
\rho_A=\rho_B= \frac{\left(\sum_{i=1}^k U_i\right) \left(\sum_{i=1}^k U_i\right)^\dag}{\mathrm{tr} \left[\left(\sum_{i=1}^k U_i\right) \left(\sum_{i=1}^k U_i\right)^\dag\right]}.
\eeq
This expression is similar to \eqref{Ginrho} with $X$ replaced by $\sum_{i=1}^k U_i$. Again the Schmidt coefficients $\{\lambda_i\}$ are given by the eigenvalues of the reduced density matrix $\rho_A$ or $\rho_B$, equivalently the unnormalised Schmidt coefficients are given by the square singular values of $\sum_{i=1}^k U_i$. The derivation of the spectrum of this matrix is more involved since it is a sum of independent random matrices. We refer the reader to \cite{random-density} for a detailed derivation, where it is shown that for large $N$ and $k\geq 2$ the eigenvalue density is given by $\rho_k(\lambda,N) \!  =\!  N \mu_k(N \lambda)$ with
\bea
\mu_k(x) =  \frac{\sqrt{4k(k-1)x -k^2 x^2}}{2\pi (k x-x^2)}, \,\,\, x\!\in\! [0,4 \tfrac{k-1}{k}]. \label{densityk}
\eea
For $k=1$ one cannot use the eigenvalue density since the spectrum collapses to a zero. Instead, from \eqref{krho} we see that for $k=1$ ones has $\rho_A=\rho_B=\mathbb{I}/N$ which is precisely the maximally entangled state as claimed above. Furthermore, for $k\to\infty$, the eigenvalue density \eqref{densityk} reduces to the Marchenko-Pastur distribution \eqref{density}, which corresponds to the HS ensemble.

To obtain the expected minimal target value under the structured ensemble we use an analogous expression of \eqref{avgA2}, where the Schmidt coefficients are now distributed as the eigenvalues of \eqref{krho}. For large $N$ we can again approximate the last term in \eqref{avgA2} by
\bea
\frac{1}{N^{2}}\!\!\expec{\sum_k \sum_l N\sqrt{\lambda_k \lambda_l}}_{\!\!\! k} \!\!  \approx \!\!\! \int_0^\infty \!\!\!\! \int_0^\infty \!\!\! \!\! \!\! \mu_k(x)\mu_k(y)\! \sqrt{xy} dx dy.
\eea
Substituting $\mu_k(x) $ from \eqref{densityk} yields
\bea
C_k\!\! &:=& \!\! \int_0^\infty \! \!\!  \int_0^\infty \!\!\!  \mu_k(x)\mu_k(y)\! \sqrt{xy} dx dy  \nonumber \\
\!\!&=&\!\! \frac{k}{\pi^2}  \left(\! 2\sqrt{k-1}\!-\!(k-2) \arcsin\left(\frac{2 \sqrt{k-1}}{k}\right)\! \right)^2 \label{tmpxx}
\eea
Plugging this as well as \eqref{tmpxx2} into the analogous expression of \eqref{avgA2}, we obtain for $k\geq2$,
\beq
\expec{\mathcal{A}_\infty}_k = 2-\frac{16 G }{\pi^2} C_k \label{reskens}
\eeq
with $C_k$ given in \eqref{tmpxx}.

\section{Discussion}
In this letter we obtain the expected quantum violations of the Bell inequality \eqref{ourBell} for $N$-dimensional random pure states, assuming the conjectured best measurement operators. The results, as summarised in Figure \ref{fig1}, show that for small values of $N$ the Bell inequality is not violated on average, while for large values it is. This holds for both the Hilbert-Schmidt (HS) ensemble, as well as structured ensembles with parameter $k$. Using techniques from random matrix theory (RMT), in particular the relation to fixed-trace ensembles, we obtain the expected minimal target value for the HS ensemble analytically for $N=2$ and $N\to\infty$ and numerically for an intermediate range of $N$. For the structured ensemble we arrive at an analytical expression for $N\to\infty$ and arbitrary $k\geq 2$, while we analyse it numerically for an intermediate range of $N$ and fixed values $k=2,3,6,12$. Most interestingly, even in the structureless case, i.e.\ the HS ensemble, for large $N$, one almost surely violates the Bell inequality. This is further explained in Appendix \ref{appA}, where we numerically look at higher moments of the minimal target value in the HS ensemble and, using a relation to the $O(n)$ model with $n=-2$, also indicate how to analytically obtain the finite $N$ results together with the higher moments. Note that the RMT techniques employed here are very general and one can foresee several future extensions, such as the Bell setting with $2$ parties, $M\geq2$ measurements and $N$ outcomes, as well as other cases which involve Hilbert spaces of large dimensions. Let us finally comment on experimental implementations. Recent experiments already analyse the quantum violations of the Bell inequality \eqref{ourBell} for $N=3$ \cite{exp}. When moving to larger $N$ and having noise in the preparation of the state, the results presented here become potentially relevant. In this context, we recall that the HS ensemble describes unstructured noise, which can be seen as a worst case scenario in terms of violations of Bell inequalities for any real noise in an experimental setting. As seen explicitly, structured noise leads to larger violations of the Bell inequalities.

\acknowledgments
The authors thank G.~Akemann for discussion and comments. MA is supported by the European Research Council under the European Union's Seventh Framework Programme (FP/2007/2013)/ ERC Grant Agreement n. 307074. S.Z.~acknowledges support by Nokia Technologies, Lockheed Martin and the University of Oxford through the Quantum Optimisation and Machine Learning (QuOpaL) Project. The early stages of this work were support by CNPq (Grant 307700/2012-7) and PUC-Rio.

\appendix

\section{On finite $N$ corrections and the relation to the $O(n)$ model} \label{appA}

\begin{figure}
\begin{center}
\includegraphics[width=3.2in]{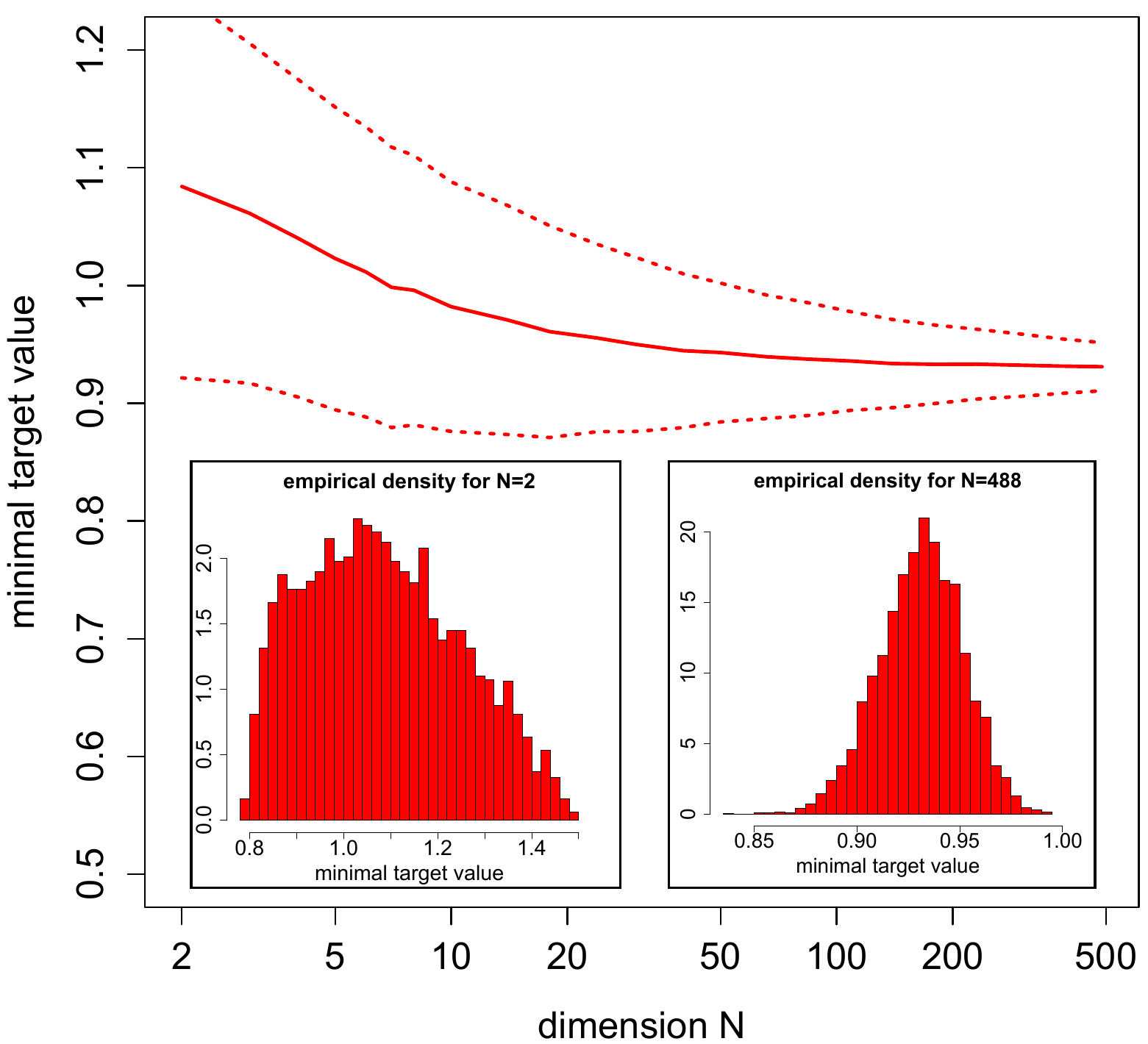}
\caption{Mean minimal value of $\mathcal{A}_N$ (solid line) plus minus its standard deviation (dashed lines) under the measure of random pure states from the HS ensemble. The inlays show the histograms from $N=2$ and $N=488$.}
\label{fig2}
\end{center}
\end{figure}

In this appendix we briefly sketch an approach for computing $\expecr{\mathcal{A}_N}$ at finite $N$ together with the $k$-th moments $\expecr{\mathcal{A}^k_N}$, focusing on the HS ensemble. Consider the $k$-th moment $\expecr{\mathcal{A}^k_N}$. From \eqref{target} it takes the form,
\beq
\expecr{\mathcal{A}^k_N} \!\! = \!\!\! \!\!\! \sum_{i_1,i_2,\ldots,i_{2k}}  \prod^k_{j=1}M_{i_{2j-1},i_{2j}} \expec{\prod^{k}_{j=1} \lambda_{i_{2j-1}}^{\frac{1}{2}} \lambda_{i_{2j}}^{\frac{1}{2}}}_{\!\!\!\mathrm{HS}}\!\!\!\!.
\eeq
It is instructive to first look at the second moment for definiteness. The previous expression shows that to leading order in $N$ ones has $\expecr{\mathcal{A}^2_N} = \expecr{\mathcal{A}_N}\expecr{\mathcal{A}_N}+\ldots$ and that thus the variance goes to zero as $N$ gets large. This can also be seen from the numerical data shown in Figure \ref{fig2}, where apart from the mean target value we also display its standard deviation given through $(\expecr{\mathcal{A}^2_N}- \expecr{\mathcal{A}_N}^2)^{1/2}$. Furthermore, we also show the histograms of the empirical density for $N=2$ and $N=488$. The latter confirms that for large $N$ almost all random pure states from the HS ensemble violate the inequality.

We now describe the analytical approach which can be used to derive $\expecr{\mathcal{A}_N}$ at finite $N$ together with its $k$-th moments.
Let $\mathcal{P}_r$ denote the set of partitions of the set ${1,2,\ldots,r}$ and $\sim$ the associated equivalence relation. Then we may write,
\bea
\expecr{\mathcal{A}^k_N}& =& \sum_{p \in \mathcal{P}_{2k}} \left(\sum_{(i_1,\ldots,i_{2k}) \in \Omega_p}\prod^k_{j=1}M_{i_{2j-1},i_{2j}} \right) \times \nonumber \\
&& \times \expecrr{\prod^{|p|}_{i=1} \lambda_i^{\frac{|p_i|}{2}} }.
\eea
Here, $\Omega_p = \{(i_1,\ldots,i_{2k}) : i_l = i_{l'} \iff l \sim l'  \}$ and $p_i$ is $i$-th part of the partition $p$. Note that we have used the invariance of \eqref{jpdf} under permutations to rewrite the product of eigenvalues. By using \eqref{deltaLUE} and making the change of variables $\lambda_i = \zeta_i^2$ we have, 
\beq
\expecrr{\prod^{|p|}_{i=1} \lambda_i^{\frac{|p_i|}{2}} } = \frac{\Gamma(N^2)}{\Gamma(N^2 + k)}N^k
\expecO{\prod^{|p|}_{i=1} \zeta_i^{|p_i|} },
\eeq
where the second expectation value is computed in the $O(-2)$ model with the restriction that $\zeta_i >0$. The general $O(n)$ model has been studied in \cite{Onmodel1,Onmodel2} where it was shown that all products of resolvents could be computed in a recursive scheme known as topological recursion. We can place our problem within this framework by again using the invariance under permutation to rewrite,
\beq
\expecO{\prod^{|p|}_{i=1} \zeta_i^{|p_i|}} \!\! = \! \frac{(N-|p|)!}{N!}\expecO{\sum_{j_1 \neq j_2 \neq \ldots \neq j_{|p|}}\prod^{|p|}_{i=1} \zeta_{j_i}^{|p_i|}}\!\!\!.
\eeq
By a standard inclusion-exclusion argument one can write the right hand side as products of traces. For example, if $|p|=3$, we would have,
\bea
&&\expecO{\sum_{j_1 \neq j_2 \neq j_3} \zeta_{j_1}^{|p_1|}\zeta_{j_2}^{|p_2|}\zeta_{j_3}^{|p_3|}} \nonumber\\
&=& \!\!\expecO{\sum^N_{j_1, j_2, j_3 = 1} \zeta_{j_1}^{|p_1|}\zeta_{j_2}^{|p_2|}\zeta_{j_3}^{|p_3|}} \!\!- \expecO{\sum^N_{j_1, j_3 = 1} \zeta_{j_1}^{|p_1| + |p_2|}\zeta_{j_3}^{|p_3|}} \nonumber \\
&&-\expecO{\sum^N_{j_2, j_3 = 1} \zeta_{j_2}^{|p_2|}\zeta_{j_3}^{|p_1|+|p_3|}}\!\! -\expecO{\sum^N_{j_1, j_2 = 1} \zeta_{j_1}^{|p_1|}\zeta_{j_2}^{|p_2|+|p_3|}} \nonumber\\
&&+ 2\expecO{\sum^N_{j_1 = 1} \zeta_{j_1}^{|p_1|+|p_2|+|p_3|}}. 
\eea
Finally we note that products of traces can be obtained directly from the asymptotic behaviour near infinity of products of resolvents. Details of this calculation will be presented elsewhere.

%
%

\section{Numerical analysis}\label{appB}

The numerical analysis for the HS ensemble is based on \eqref{Ginrho}. For a fixed value of $N$ we sample $i.i.d.$ normal random variables as entries of the $N\times N$ complex Wishart matrix $X$ and then numerically calculate the eigenvalues of $X X^\dag/\mathrm{tr} (X X^\dag)$. From this we obtain a sample of $\{\la_i\}$ which is used to compute \eqref{target}. Repeating this 1000 times and taking the sample average yields an estimate for $\expecr{\mathcal{A}_N}$. This procedure is done for $N$ in the range of $2$ to around $500$ in exponentially increasing increments, resulting in the data points in Figure \ref{fig1}. It is checked that for $N=2$ and $N=500$ the values agree with the analytical predictions \eqref{A2} and \eqref{Ainf} respectively. Furthermore, for $N=100$ we also use the 1000 samples of $\{\la_i\}$ to numerically check the functional form of the eigenvalue density \eqref{density}. The above numerical analysis centred around \eqref{Ginrho} is particularly useful to obtain numerical estimates of $\expecr{\mathcal{A}_N}$ for intermediate values of $N$, alternatively for large $N$ one can also simulate the Coulomb gas \eqref{jpdf} using the Metropolis algorithm.  

In analogy to the HS ensemble, the numerical analysis for the structured ensemble is based on \eqref{krho}. Instead of sampling a complex Wishart matrix $X$, we have to sample $k$ independent unitary matrices and calculate $\sum_{i=1}^k U_i$. From there onwards the analysis proceeds as in the case of the HS ensemble, simply replacing $X$ by $\sum_{i=1}^k U_i$ in the remainder. To sample the $k$ independent unitary matrices we follow the procedure introduced in \cite{Francesco} (see also \cite{RUM}), which constructs a unitary matrix $U_i$ distributed according to the Haar measure from a complex Wishart matrix $X_i$ via a $QR$-decomposition. Note that an explicit Python implementation of the subroutine for sampling random unitary matrices can be found for example in \cite{Francesco}.

\end{document}